\documentstyle[12pt]{article}
\normalbaselines
\begin{document}
\bibliographystyle{unsrt}

\vbox {\vspace{5mm}}

\begin{center}
{\huge\bf Selection of Squeezed States   }\\[8mm]
{\huge\bf via Decoherence}\\[15mm]

Gh.--S. Paraoanu$^{*}$
 \\
{\it Department of Physics, University of Illinois at Urbana-Champaign, \\ 1110 W. Green St., Urbana, IL 61801, USA}\\[2mm]
 
\end{center}

\vspace{2mm}

\begin{abstract}

In the framework of Lindblad theory for open quantum systems, we 
calculate the entropy of a damped quantum harmonic oscillator which is 
initially in a quasi-free state. The maximally predictable states are identified as those states producing the minimum 
entropy increase after a long enough time. In general, the states with a squeezing parameter depending on the environment's diffusion coefficients and friction constant are singled out, but if the friction constant is much smaller than the oscillator's frequency, coherent states (or thermalized coherent states) are obtained as the preferred classical states.

\end{abstract}
\vspace{1cm}

\newpage

~

\vspace{1cm}

Although quantum theory has reported during its almost one century of existence spectacular successes in explaining phenomena at the atomic scale and below, the relation between quantum and classical reality is still a problem not very well understood. At
 the heart of this problem is the superposition principle, which gives counter-intuitive results when applied to macroscopic bodies (the infamous Schr\"{o}dinger cat is probably still the best illustration of this apparent paradox). 

In the last 15-16 years, the theory of decoherence \cite{1,2} has been advanced as a solution to this conundrum: macroscopic bodies are open systems, interacting continuously with their environment, and it is exactly this interaction which destroys the annoying superposition terms, thus creating superselection sectors in the Hilbert space. 

This strategy has been applied by W. H. Zurek and collaborators \cite{zc,z} to study the decoherence mechanism in the case of a quantum harmonic oscillator undergoing quantum Brownian motion \cite{leggett}. In this model, the system is coupled linearly through its coordinate to a heat bath of harmonic oscillators which constitute the environment. In their seminal paper \cite{z}, Zurek, Habib and Paz [ZHP] showed that, if one adopts ``predictability sieve'' as a criterion for classicality - so that the preferred classical states are those leading to a minimum increase in entropy - then coherent states are singled out as the maximally predictable pure states.

Recently, M. R. Gallis \cite{gallis} applied ZHP's criterion to a harmonic oscillator weakly coupled to an environment; the oscillator  is treated as an open quantum system in the framework of Lindblad theory \cite{lin}. The theory of this system was initially developed by Sandulescu and Scutaru \cite{scut}
 and further continued in \cite{isar}. One of the advantages  of adopting this framework is that the Lindblad form for the evolution operator is a rather  general phenomenological master equation which preserves the positivity of the density matrix and  entropy and which encompasses a large variety of concrete physical problems, from quantum optics to nuclear physics.

In this paper we generalize Gallis' result (for Lindblad generators linear in position and momentum) to the case of an arbitrary large coupling between the system and the environment and show that for a damped harmonic oscillator the preferred states have a squeezing parameter depending on the environment's friction constant and diffusion coefficients. The calculation is done exactly, not perturbatively 
as in \cite{gallis}, which allow us to consider as a particular case the weak-coupling limit in which Gallis worked. We prove that coherent states (or thermalized coherent states, if one starts with mixed states) are indeed selected when the friction constant is much smaller than the angular velocity of the oscillator, the magnitude of the diffusion coefficients being irrelevant.

The general form of a completely dissipative mapping L is given by the Lindblad theorem \cite{lin}
\begin{equation}
L(\rho)=-\frac{i}{\hbar}[H,\rho]+\frac{1}{2\hbar}
\sum_{j}([V_{j}
\rho,V_{j}^{+}]+[V_{j},\rho V_{j}^{+}]),
\end{equation} 
so the evolution equation for an open quantum system is
\begin{equation}
\frac{d\rho (t)}{dt}=L(\rho (t)). \label{ev}
\end{equation}
For a harmonic oscillator, $H=\frac{p^{2}}{2m}+\frac{m\omega^{2}}{2}q^{2}$.
To get an exact soluble model, as shown in \cite{scut}, one has to take $V_{j}$ as a first-degree non-commutative polynomial in coordinate and momentum, and because $p$ and $q$ span the linear space of the first-degree polynomials in $p$ and $q$, there exist only two linear independent operators $V_{1}$ and $V_{2}$, with $V_{i}=a_{i}p+b_{i}q$, $i=1,2$ and $a_{i}$, $b_{i}$ complex numbers. 
With the notations 
\begin{equation}
D_{qq}={\hbar\over2}\sum_{j=1}^{2}|a_{j}|^{2},~~D_{pp}={\hbar\over2}\sum
_{j=1}^{2}|b_{j}|^{2},~~D_{pq}=D_{qp}=-{\hbar\over2}Re\sum_{j=1}^{2}
a_{j}^{*}b_{j},
\end{equation}
for the entries of the phase-space diffusion matrix ($D_{pp}$, $D_{pq}$ and $D_{qq}$ are called diffusion coefficients) and   
\begin{equation}
\lambda=-Im\sum_{j=1}^{2}a_{j}^{*}b_{j},
\end{equation}
representing the friction constant, $L(\rho )$ takes the form
\begin{eqnarray}
L(\rho )&=&-{i\over\hbar}[H,\rho ]
-{i\lambda\over 2\hbar}[q,\rho p+p\rho ]+{i\lambda\over 2\hbar}[p,\rho q+q\rho ]- \nonumber\\ & &
-{D_{qq}
\over\hbar^2}[p,[p,\rho ]]
-{D_{pp}\over\hbar^2}[q,[q,\rho ]] 
+{2D_{pq}\over
\hbar^2}[p,[q,\rho ]]
.\label{d} \end{eqnarray}
The evolution equation (\ref{ev}) with the dissipative mapping $L$ given by (\ref{d}) has the property that the set of quasi-free states \cite{ss} is invariant under the action of the evolution mapping $\rho (0)\rightarrow \rho (t)$. 
Quasi-free states are generalizations of the coherent and squeezed states, and are completely characterized by the correlation functions:
\begin{eqnarray}
\sigma_{qq}&=&Tr(\rho q^{2})-[Tr(\rho q)]^{2},\\ 
\sigma_{pp}&=&Tr(\rho p^{2})-[Tr(\rho p)]^{2},\\
\sigma_{pq}&=&Tr(\rho\frac{pq+qp}{2})-Tr(\rho p)Tr(\rho q).
\end{eqnarray}
Let us denote the correlation matrix by
\begin{equation}\sigma
=\left(\matrix{m \omega \sigma_{qq}&\sigma_{pq}\cr\sigma_{pq}&
{\sigma_{pp}\over m \omega}\cr}\right).
\end{equation}
Because of the Heisenberg inequality, $\sigma_{qq}\sigma_{pp}-\sigma_{pq}^{2}=det\sigma\geq\frac{\hbar^{2}}{4}$, $\sigma$ is a complete positive matrix which can be diagonalized \cite{bal}
\begin{equation}
\sigma ={\hbar A \over 2}O^{T}\left(\matrix{\aleph^{2}&0\cr 0&
\aleph^{-2}}\right)O,\label{dec}
\end{equation}
where 
\begin{equation}
A=\frac{2}{\hbar}\sqrt{det \sigma}\geq 1 \label{A}\end{equation}
is the ``area'' in phase space measured in units of $\frac{\hbar}{2}$; O is an orthogonal symplectic matrix,
\[ O=\left(\matrix{\cos\theta&-\sin\theta\cr\sin\theta&
\cos\theta\cr}\right),
\]
and $\aleph$ is a positive real number (the squeezing parameter of the state).
The degree of purity is indicated by $A\geq 1$; pure states for example correspond to  $A=1$. The matrix $O$ describes a rotation in phase space, and $\aleph $  a change of scale (squeezing and dilatation) in the phase-space coordinates $p$ and $q$. Given a quasi-free state, $A$ is uniquely determined, but $\theta$ and $\aleph$ are determined only up to the transformations
\begin{equation}
\theta\rightarrow n\pi + \theta , ~~~~(n~ {\rm integer})
\end{equation}
and
\begin{equation}
\begin{array}{c} \aleph\rightarrow\aleph^{-1}; \\ \theta\rightarrow\frac{\pi}{2}+\theta .
\end{array}\label{tran}
\end{equation}
The statistical entropy of a quasi-free state is \cite{ag}
\begin{equation}
S=\frac{A+1}{2}\ln\frac{A+1}{2}-\frac{A-1}{2}\ln\frac{A-1}{2}. \label{entr}
\end{equation}
The study of the time-evolution of $S$ is equivalent to the study of the evolution of $A$, {\it i.e.} how $det\sigma (t)$ changes in time. What we will do next is to calculate $det\sigma (t)$ at $t\gg \lambda^{-1}$, and find the state $\sigma (0)$ for which $S(t)$ is minimum (because $S$ increases with $A$, this will be equivalent with $det\sigma (t)$ being minimum).

The time evolution for the correlations $\sigma_{pp}$, $\sigma_{qq}$ and $\sigma_{pq}$ is known (see Eq. (4.55) from \cite{scut})
\begin{equation}
\sigma (t)={\cal R}(t)(\sigma (0) -\sigma (\infty )){\cal R}^{T}(t)+\sigma (\infty ), \label{sigma}
\end{equation}
with 
\begin{equation}
{\cal R}(t)= \left(\matrix{\delta (t)&-m\omega\beta (t)\cr -\frac{\gamma (t)}{m\omega }&\alpha (t)}\right)
\end{equation}
(see Eq. (4.51), \cite{scut}).
Now, (4.22) and (4.23) from \cite{scut} imply that the time evolution for $\alpha ,\beta ,\gamma $ and $\delta $ is given by
\begin{equation}
\left(\matrix{\alpha (t)&\beta (t) \cr \gamma (t) &\delta (t)}\right)=e^{-\lambda t}\left(\matrix{\cos\omega t&-\frac{1}{m\omega}\sin\omega t\cr m\omega \sin\omega t&\cos\omega t}\right).
\end{equation}
We have then 
\begin{equation}
{\cal R}(t)=e^{-\lambda t}\left(\matrix{\cos\omega t&\sin\omega t\cr -\sin\omega t&\cos\omega t}\right),
\end{equation}
and using (\ref{sigma}) and the notations
\begin{eqnarray}
{\cal T}_{pp}(t)&=&\frac{\sigma_{pp}(\infty )}{m\omega}\cos^{2}\omega t + m\omega\sigma_{qq}(\infty )\sin^{2}\omega t + 2\sigma_{pq}(\infty )\sin\omega t\cos\omega t \label{oho1}\\
{\cal T}_{qq}(t)&=&\frac{\sigma_{pp}(\infty )}{m\omega}\sin^{2}\omega t+m\omega\sigma_{qq}(\infty )\cos^{2}\omega t-2\sigma_{pq}(\infty )\sin\omega t\cos\omega t \label{oho2}\\
{\cal T}_{pq}(t)&=& \left[ m\omega\sigma_{qq}(\infty )-
\frac{\sigma_{pp}(\infty )}{m\omega }\right]\sin\omega t\cos\omega t + \sigma_{pq}(\infty )(\cos^{2}\omega t -\sin^{2}\omega t)\label{oho3}
\end{eqnarray}
we finally get
\begin{eqnarray}
det\sigma (t)&=&e^{-4\lambda t}det[\sigma (0)-\sigma (\infty )]+  \nonumber \\
 & &+e^{-2\lambda t}\left[ m\omega (\sigma_{qq}(0)-\sigma_{qq}(\infty )){\cal T}_{pp}(t)+\frac{\sigma_{pp}(0)-\sigma_{pp}(\infty )}{m\omega}{\cal T}_{qq}(t)-2(\sigma_{pq}(0)-\sigma_{pq}(\infty )){\cal T}_{pq}(t) \right]+ \nonumber \\
 & &+det\sigma (\infty).\nonumber
\end{eqnarray} 
So, what we have here is an exact expression for the phase-space area ``occupied'' by the system at any time t -- and implicitly for the entropy $S$, via the relations (\ref{A}) and (\ref{entr}).

As emphasized by Gallis \cite{gallis}, classicality should be an enduring property and not a transient one, so large-time scales should be relevant; in our case, the time scale is set by $\lambda^{-1}$. Thus, for $t\gg \lambda^{-1}$, the term $e^{-4\lambda t}det[\sigma (0)-\sigma (\infty )]$ in the expression of $det\sigma (t)$ can be neglected, and to get the maximal predictable states we have to minimize the quantity
\begin{equation} 
e^{-2\lambda t}\left[ m\omega (\sigma_{qq}(0)-\sigma_{qq}(\infty )){\cal T}_{pp}(t)+\frac{\sigma_{pp}(0)-\sigma_{pp}(\infty )}{m\omega}{\cal T}_{qq}(t)-2(\sigma_{pq}(0)-\sigma_{pq}(\infty )){\cal T}_{pq}(t) \right].\label{x}
\end{equation}
Using the decomposition (\ref{dec}) and retaining only the part dependent on the initial state parameters in (\ref{x}), we find the expression:
\begin{eqnarray}
\frac{\hbar A(0)}{2}\{ \cos^{2}\theta (0)\left[\aleph^{2}(0){\cal T}_{pp}(t)+\aleph^{-2}(0){\cal T}_{qq}(t)\right]+\sin^{2}\theta (0)\left[\aleph^{-2}(0){\cal T}_{pp}(t)+\aleph^{2}(0){\cal T}_{qq}(t)\right]- \nonumber \\
 -2\sin\theta (0)\cos\theta (0)\left[\aleph^{-2}(0)-\aleph^{2}(0)\right]{\cal T}_{pq}(t)\} \nonumber,
\end{eqnarray}
which, when minimized with respect to $\aleph (0)$ and $\theta (0)$ gives
\begin{equation}
\tan 2\theta^{*}(0)=\frac{2{\cal T}_{pq}}{{\cal T}_{pp}-{\cal T}_{qq}}\label{teta}
\end{equation}
and
\begin{equation}
\aleph^{*}(0)=\left[\frac{\mp\sqrt{({\cal T}_{pp}-{\cal T}_{qq})^{2}+4{\cal T}_{pq}^{2}}+{\cal T}_{pp}+{\cal T}_{qq}}{\pm\sqrt{({\cal T}_{pp}-{\cal T}_{qq})^{2}+4{\cal T}_{pq}^{2}}+{\cal T}_{pp}+{\cal T}_{qq}}\right]^{1/4}\label{hi}
\end{equation}
(the two values for $\aleph^{*}(0)$ represent the same state, via (\ref{tran}) and the observation that $\tan (2\theta )$ is invariant under $\theta\rightarrow \frac{\pi}{2}+\theta $). 

Now, $\sigma_{pp}(\infty )$, $\sigma_{qq}(\infty )$ and $\sigma_{pq}(\infty )$ can be expressed in terms of the diffusion coefficients and the friction constant (see equation (3.53) in \cite{scut}) in the form
\begin{eqnarray}
\sigma_{qq}(\infty )&=&\frac{1}{2(m\omega )^{2}\lambda (\lambda^{2}+\omega^{2})}\left[ (m\omega )^{2}(2\lambda ^{2}+\omega^{2})D_{qq}+\omega^{2}D_{pp}+2m\omega^{2}\lambda D_{pq}\right]\\
\sigma_{pp}(\infty )&=&\frac{1}{2\lambda (\lambda^{2}+\omega^{2})}\left[(m\omega )^{2}\omega^{2}D_{qq}+(2\lambda^{2}+\omega^{2})D_{pp}-2m\omega^{2}\lambda D_{pq}\right]\\
\sigma_{pq}(\infty )&=&\frac{1}{2m\lambda (\lambda^{2}+\omega^{2})}\left[ -\lambda (m\omega )^{2}D_{qq}+\lambda D_{pp}+2m\lambda^{2}D_{pq}\right]
\end{eqnarray}
Inserting these expressions into (\ref{oho1}) - (\ref{oho3}) and further into (\ref{hi}), we obtain
\begin{equation}
\aleph^{*}(0)=\left[\frac{\mp\sqrt{\left(m\omega D_{qq}-\frac{D_{pp}}{m\omega}\right)^{2}+4D_{pq}^{2}}+\sqrt{1+\frac{\omega^{2}}{\lambda^{2}}}\left(m\omega D_{qq}+\frac{D_{pp}}{m\omega}\right)}{\pm\sqrt{\left(m\omega D_{qq}-\frac{D_{pp}}{m\omega}\right)^{
2}+4D_{pq}^{2}}+\sqrt{1+\frac{\omega^{2}}{\lambda^{2}}}\left(m\omega D_{qq}+\frac{D_{pp}}{m\omega}\right)}\right]^{1/4}\label{sq} 
\end{equation}

This is the general form for the squeezing parameter of the quasi-free states selected via ZHP's ``predictability sieve''. In particular, for $\lambda\ll\omega$, the squeezing parameter becomes $\aleph^{*}(0)=1$. This basically means that if the initial states are pure then coherent states will be selected. So, the same 
result as in \cite{z} and \cite{gallis} is obtained in the limit of a small friction constant (we do not have to restrict the discussion, as in \cite{gallis}, to the case in which all the effects of the environment are relatively small comparing to the Hamiltonian -- the diffusion coefficients can be arbitrary large). 

Another interesting particular case in which coherent states are the preferred states, as can be seen from (\ref{sq}), is that of an isotropic diffusion in phase space ($\frac{D_{pp}}{m\omega}=m\omega D_{qq}$ and $D_{pq}=0$), a model which was previously studied in the framework of Lindblad theory by Ingarden and Kossakowski \cite{ing}. 

These results are  also independent of the degree of purity of the initial state. If, instead of starting with a pure state, the system is initially in a mixed state ($A(0)>1$), then states with squeezing parameter given by (\ref{sq}) will be singled out. These states are thermalized squeezed states in general, and become thermalized coherent state in the limit $\lambda\ll\omega$ or for  $\frac{D_{pp}}{m\omega}=m\omega D_{qq}$ and $D_{pq}=0$.

\vspace{1cm}

The author gratefully thanks H. Scutaru for enlightening discussions and invaluable ideas.

\vspace{3cm}

$^{*}$ On leave  from the {\it Department of Theoretical Physics, Institute of Physics and Nuclear Engineering ``Horia Hulubei'', Bucharest-Magurele, PO Box MG-6, Romania}. Electronic address: paraoanu@physics.uiuc.edu  .


\begin{thebibliography}{99}

\bibitem{1}
W. H. Zurek, {\it Phys. Rev. D} {\bf 24}, 1516 (1981); {\bf 26}, 1862 (1982).

\bibitem{2} 
E. Joos and H. D. Zeh, {\it Z. Phys. B} {\bf 59}, 223 (1985).

\bibitem{zc}
W. G. Unruh and W. H. Zurek, {\it Phys. Rev. D} {\bf 40}, 1071 (1989);
B. L. Hu, J. P. Paz and Y. Zhang, {\it Phys. Rev. D} {\bf 45}, 2843 (1992);
J. P. Paz, S. Habib and W. H. Zurek, {\it Phys. Rev. D} {\bf 47}, 488 (1993).

\bibitem{z}
W. H. Zurek, S. Habib and J. P. Paz, {\it Phys. Rev. Lett.} {\bf 70}, 1187
(1993).

\bibitem{leggett}
A. O. Caldeira and A. J. Leggett, {\it Physica (Amsterdam)} {\bf
121A}, 587 (1983);
A. O. Caldeira and A. J. Leggett, {\it Phys. Rev. A} {\bf 31}, 1057 (1985).

\bibitem{gallis}
M. R. Gallis, {\it Phys. Rev. A} {\bf 53}, 655 (1996).

\bibitem{lin}
G. Lindblad, {\it Commun. Math. Phys.} {\bf 48}, 119 (1976).

\bibitem{scut} 
A. Sandulescu and H. Scutaru, {\it Ann. Phys. (N.Y.)} {\bf 173}, 277 (1987).

\bibitem{isar}
A. Isar, W. Scheid and A. Sandulescu, {\it J. Math. Phys.} {\bf 32}, 2128 (1991);
A. Isar, A. Sandulescu and W. Scheid, {\it ibid.} {\bf 34}, 3887 (1993);
A. Isar, {\it Helv. Phys. Acta} {\bf 67}, 436 (1994);
Gh.-S. Paraoanu and H. Scutaru, Phys. Lett. A {\bf 238}, 219 (1998).

\bibitem{ss}
H. Scutaru, {\it Phys. Lett. A} {\bf 200}, 91 (1995).

\bibitem{bal}
R. Balian, C. De Dominicis and C. Itzykson, {\it Nucl. Phys.} {\bf 67}, 609 (1965).

\bibitem{ag}
G. S. Agarwal, {\it Phys. Rev. A} {\bf 3}, 828 (1971).

\bibitem{ing}
R. S. Ingarden and A. Kossakowski, {\it Ann. Phys.} (N.Y.) {\bf 89}, 451
(1975).




\end{thebibliography}
\end{document}